\documentclass[11pt,twoside]{article}

\usepackage{asp2004}
\usepackage{epsf}
\usepackage{psfig}
\usepackage{lscape}
\usepackage{graphicx}
\usepackage{amssymb}

\begin{document}
\title{The evolution of massive stars in the context of V838 Monocerotis}   
\author{R. Hirschi}   
\affil{Dept. of Physics and Astronomy, University of Basel, Klingelberstr. 82, 4056 Basel, Switzerland}    

\begin{abstract} 
The aim of this paper is to look at the evolution of massive stars in
order to determine whether or not the progenitor of V838 Mon may be a massive star.
In the first part of this paper, the evolution of massive stars around
solar metallicity is described, especially the evolution in the Hertzsprung-Russell (HR)
diagram. Then, using the observational constraints, the probable
progenitors (and their evolution) are described. 
Using models of single stars, no progenitor can be found 
amongst massive stars that can
satisfy all the observational constraints.
Wolf-Rayet stars (stars with initial masses above about 30 $M_\odot$, which
have lost their hydrogen rich envelopes) could explain 10 to 100 $M_\odot$ of circumstellar
material but they are very luminous ($L \gtrsim 10^5\,L_\odot$). Main sequence
stars crossing the HR diagram and becoming red supergiants (RSG) can have very 
low effective temperatures but take
thousands of years to cross over. Be stars (fast rotating stars with a
mass around 10 $M_\odot$), which form disk or B stars accreting matter from a binary companion 
of a similar mass would need to be compared in detail with the
observational constraints. In the future, there will hopefully be 
further observational constraints on the models coming from the mass and nature (interstellar or
circumstellar) of the material producing the light echo and from a frequency estimate of 
spectacular objects such as V838 Mon.
\end{abstract}

\section{Introduction}
V838 Monocerotis is a peculiar object, which had a major outburst in 
February 2002 \citep{Kim02}. Its V magnitude first rose to 10th 
magnitude early in January. Then it rose to a peak magnitude of 6.5 around February 
2$\rm ^{nd}$. Afterward it 
had two secondary peaks to magnitudes 7 and 7.5 in March and April 
respectively. Finally a decline brought the V magnitude down to 15 by mid-May 2002.
Note that V838 became (peak around 4 or 5th magnitude) and stayed brighter in 
redder bands (for example the J and K bands). This means that the star after the 
eruption became a very cool star (T$\rm _{eff} \lesssim 2000$ K, 
see contribution by Martin in this volume). 
Similar objects are M31 RV and V4332 Sgr 
(Kimeswenger in this volume).
Apart from its outbursts, V838 Mon is famous for the light echoes observed with 
Hubble Space Telescope 
(Bond in this volume). Indeed light emitted during the bursts is 
scattered by the surrounding material towards us with a delay. At a given time, 
we receive light from an
ellipsoidal surface of the circumstellar material (Sugerman in this volume).
Ashok (in this volume) with Spitzer observations of dust thermal re-radiation 
estimates the amount of circumstellar material to be between 10 and 100 $M_\odot$.
It is however not known yet whether this material is of interstellar origin or of circumstellar 
origin (lost by V838 progenitor).
SiO maser emission was also detected in V838 Mon 3 years
after its outburst (Deguchi in this volume).

There are several additional observational constraints concerning V838 and its progenitor. 
The latest distance estimate using polarimetry of the light echo is around 6 kpc 
(Sparks in this volume).
The metallicity of V838 is [Fe/H]$=-0.4 \pm 0.3$ \citep{KK04} 
corresponding approximately to the metallicity of the Large Magellanic Cloud
(LMC: Z $\sim$0.008). Most elements are in solar ratio compared to iron and the chemical 
composition is compatible with unprocessed material at the galacto-centric distance of V838.
V838 has a B3V companion and a spectroscopic fit for the progenitor of V838 
is obtained with a star with a high effective temperature (T$\sim$50\,000 K) 
and with $V\sim$15.5 and $B-V\sim$0.5 \citep{MH05}. \citet{TS06} estimate the progenitor 
luminosity to be around 1000 $L_\odot$ and the maximum luminosity during outburst to be 
around 10$^6$ $L_\odot$ (using a distance of 8 kpc).
Several models have been proposed for
the progenitor of V838: 
a born-again low mass star \citep{L05}, 
a WR star \citep{MH05}, 
a planet swallowing star \citep{RZSL06}
and a binary star merger \citep{TS06}. 
\section{Stellar evolution models}
The computer model used to calculate the results presented here is described 
in \citet{ROTXI,ROTX}. 
Convective stability is determined by the 
Schwarzschild criterion. 
Overshooting is included  with an overshooting parameter of 0.1 H$_{\rm{P}}$ 
for H and He--burning cores 
and 0 otherwise, where H$_{\rm p}$ is the pressure scale height estimated at the Schwarzschild
boundary.  
The reaction rates are taken from the NACRE \citep{NACRE} compilation. 
The effective temperature of Wolf--Rayet (WR) stars is a delicate problem, since the
winds may have a non--negligible optical thickness. A correction
has been applied in the WR stages and only there \citep[see][for more details]{ROTXI}.

There are three kinds of effects induced by rotation. First, the centrifugal 
force acts again gravity and elongates isobars at the equator. 
Second, rotation enhances mass loss. Third, it induces mixing in radiative
zones. Instabilities induced by rotation
taken into account are meridional circulation 
and secular shears. 
They affect both the transport of the chemical species and of the
angular momentum \citep{MM00}.

Since mass loss rates are a key ingredient of the models for massive stars, 
let us recall the prescriptions used.
The changes of the mass loss rates, $\dot{M}$, with  
rotation are taken into account as explained in \citet{ROTVI}.
As reference mass loss rates, 
the adopted mass loss rates are the ones of \citet{Vink00,VKL01},
who account for the occurrence of bi--stability
limits, which change the wind properties and mass loss rates.
For the domain not covered by these authors
the empirical law devised by \citet{Ja88} was used.
Note that this empirical law, which presents
a discontinuity in the mass flux near the Humphreys--Davidson limit,
implicitly accounts for the mass loss rates of LBV stars. 
For the non--rotating
models, since the empirical values
for the mass loss rates are based on 
stars covering the whole range of rotational velocities, 
one must apply a reduction factor to the empirical rates to make
them correspond to the non--rotating case. Here, the same reduction factor 
was used as in \citet{ROTVII}.
During the Wolf--Rayet phase the mass loss rates by \citet{NuLa00}
were used. 
The mass loss rates during the non--WR phases depend 
on metallicity as $\dot{M} \sim (Z/Z_{\odot})^{0.5}$ \citep{VKL01}, where
$Z$ is the mass fraction of heavy elements at the surface
of the star. 

Models were calculated at four different metallicities: $Z$ = 0.004, 0.008, 0.020 and 0.040
\citep{ROTX,ROTXI}.
The initial compositions are adapted for the different
metallicities considered here.
For the heavy elements,
the same mixture as the one
used to compute the opacity tables for solar composition \citep{IR96} was adopted. 
The models have the following initial mass fractions 
for hydrogen $X$ = 0.757, 0.744, 0.705 and 0.640 
and for helium $Y$ = 0.239, 0.248, 0.275 and 0.320
at the metallicities $Z$ = 0.004, 0.008, 0.020 and 0.040 respectively.

As initial rotation, a value equal to 300 km s$^{-1}$ is considered
on the zero age main sequence (ZAMS) for all the initial masses and metallicities considered. At solar metallicity, this
initial value produces time--averaged equatorial velocities on the main sequence (MS) well in the observed range,
i.e. between 200 and 250 km s$^{-1}$. At low metallicities this initial rotational velocity corresponds also to mean values
between 200 and 250 km s$^{-1}$ on the MS phase, while at twice the solar metallicity, the mean velocity is lower, between
160 and 230 km s$^{-1}$. Presently we do not know the distribution of the
rotational velocities
at these non--solar metallicities and thus we do not know if the adopted initial velocity corresponds 
to the average observed values. 
It may be that at lower metallicities the initial velocity distribution contains a larger number of
high initial velocities \citep{MGM99}, in which case 
the effects of rotation described below would be underestimated at low metallicity.

\section{Evolution of massive stars with rotation and mass loss}

\subsection{Impact of rotation}
The main impacts of rotation on the evolution are the following. 
Rotation increases the MS lifetime with respect to non-rotating models
(up to about 40\,\%).
Rotation strongly affects the lifetimes as blue and red supergiants (RSG).
In particular in the SMC, the high observed number of RSG can only be
accounted for by rotating models \citep{ROTVII}.
Rotation increases mass loss and therefore the minimum mass for forming WR 
stars is lowered from 37 to 22
$M_\odot$ for typical rotation at solar metallicity.
Rotation also increases the WR lifetimes (on the average by a factor two).
The observed variation  with metallicity of the fractions of type Ib/Ic supernovae with respect to 
type II supernovae (or in other words of WR to O type stars) as found by \citet{PB03} is very well 
reproduced by the rotating models, while  non--rotating models predict much too low ratios. 
 
The effects of rotation on pre--supernova models are significant between
15 and 30 $M_{\sun}$ \citep{psn04a}. Indeed,
rotation increases the core sizes (and the yields of carbon and oxygen) by a factor $\sim 1.5$. 
Above 20 $M_{\sun}$, rotation may change the radius or colour of the 
supernova progenitors (blue instead of red supergiant) and
the supernova type (IIb or Ib instead of II).
Rotation affects the lower mass limits for radiative
core carbon burning, for iron core
collapse and for black hole formation.
Finally, rotating models are able to reproduce surface enrichments during the main sequence
(in particular in nitrogen) whereas non--rotating models predict surface enrichments only 
after the first dredge--up \citep{ROTV,HL002}.

\subsection{Mass loss}
\begin{figure}[!tbp]
\centering
\includegraphics[width=10cm]{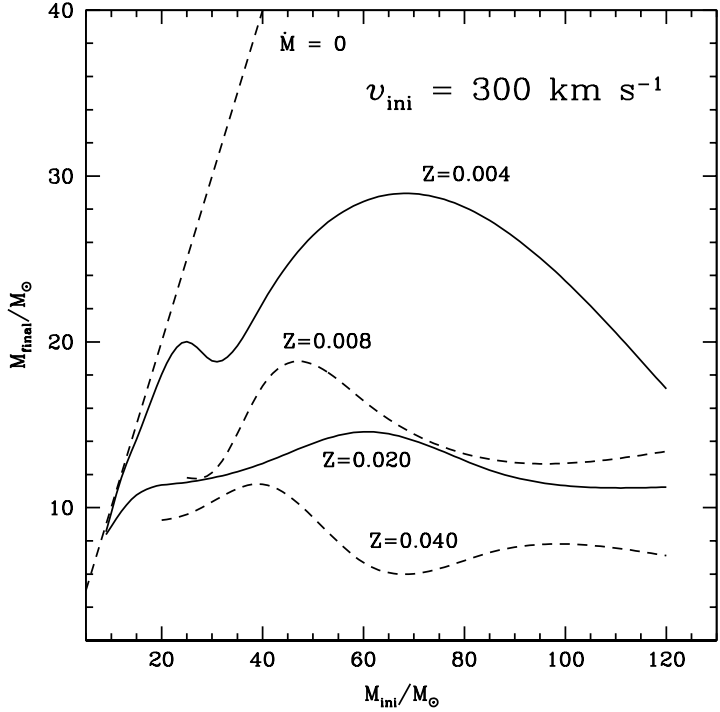}
\caption{Final mass of the rotating stellar models
as a function of the initial mass at the four metallicities $Z$ = 0.004, 0.008, 0.020 and 0.040
(from top to bottom). 
The line with slope one, labeled $\dot {\rm M}=0$, corresponding to the case without mass loss,
shows how important mass loss is for stars more massive than about 25 $M_\odot$ \citep{ROTXI}.
}
\label{mdot}
\end{figure}
Mass loss increases with metallicity (dependence on $Z^{1/2}$ in our models) 
and stellar mass (or luminosity). This of course has a direct impact on the final mass of the models, 
which is shown in Fig. \ref{mdot}. In this figure, we can see that more metal rich models end up 
with smaller final masses. Note however that models with an initial mass around 100 $M_\odot$ 
all end up with a similar final mass and they all lose about 90 $M_\odot$ during hydrogen and helium 
burnings. 
The minimum mass to form a WR star is estimated at 20, 22, 25 and 32 $M_\odot$ for metallicities 
$Z$ = 0.004, 0.008, 0.020 and 0.040 respectively.

\begin{landscape}
\begin{figure}[!tbp]
\includegraphics[angle=270, width=18cm]{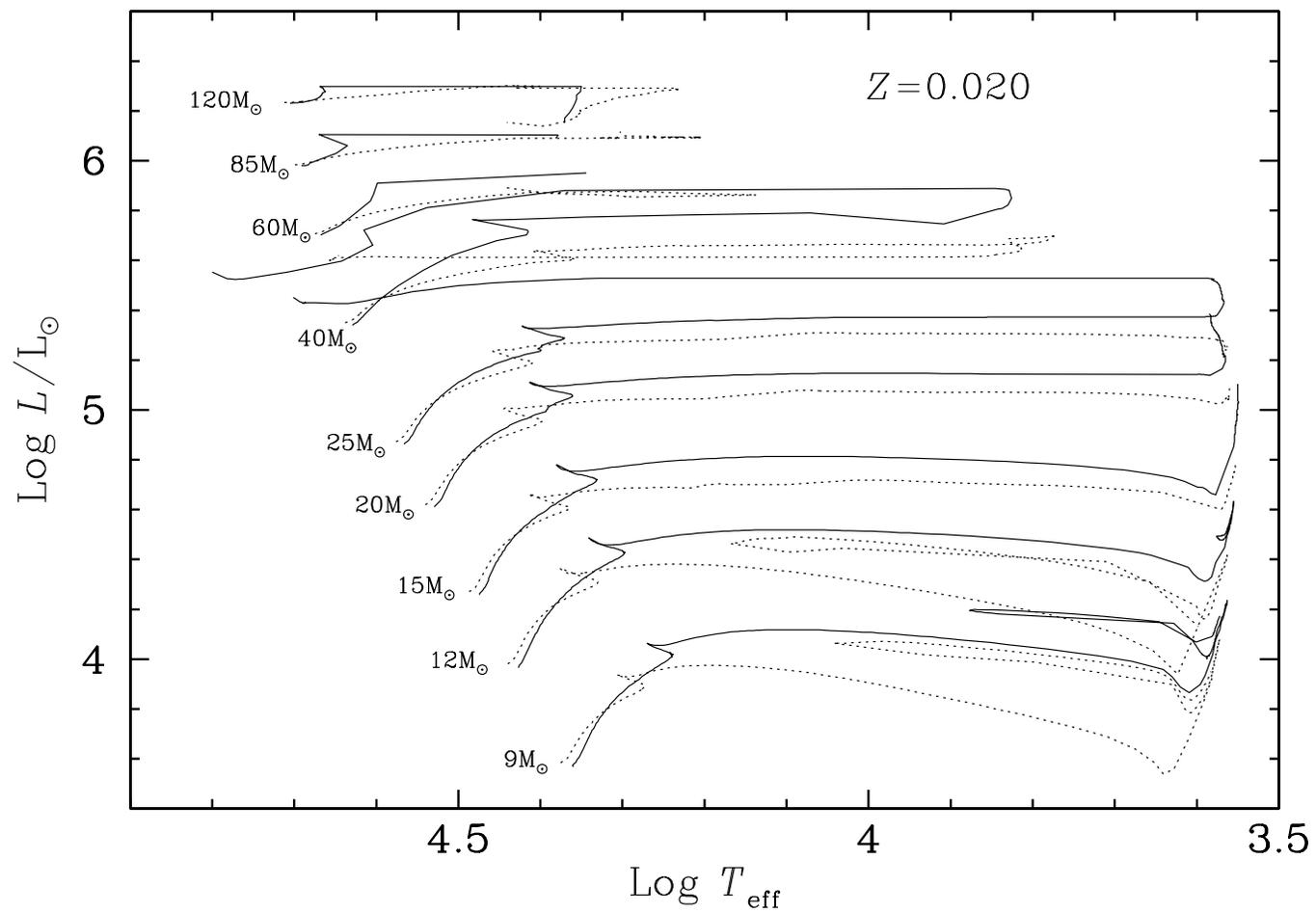}
\caption{Evolution of the $Z=0.020$ models in the HR diagram \citep{ROTX}.}
\label{zs}
\end{figure}
\end{landscape}
\subsection{Evolution in the Hertzsprung-Russell diagram}
The evolution of solar metallicity models ($Z=0.020$) in the HR diagram 
is shown in Fig. \ref{zs}. The massive star range can be divided in the following 
bins:

\begin{itemize}
\item$M\lesssim 13$ $M_\odot$: the star becomes a red supergiant (RSG) after 
the main sequence (MS) and undergoes a blue loop before ending up as a RSG.
Stars less massive than about 10 $M_\odot$ will go through the AGB phase 
(not followed in this work).
\item $M\simeq13-20$ $M_\odot$: the star crosses only once the HR diagram 
after the MS and ends as a RSG.
\item $M\simeq 20-40$ $M_\odot$: the star becomes a RSG and loses its hydrogen rich 
envelope due to mass loss and then becomes a Wolf--Rayet (WR) star.
\item $M\gtrsim 40$ $M_\odot$: the star becomes a Wolf--Rayet (WR) star and never 
goes through the RSG stage although it might go through the luminous blue variable 
(LBV) stage.
\end{itemize}

\section{Comparison between observations and models}
\begin{figure}[!tbp]
\centering
\includegraphics[width=7cm]{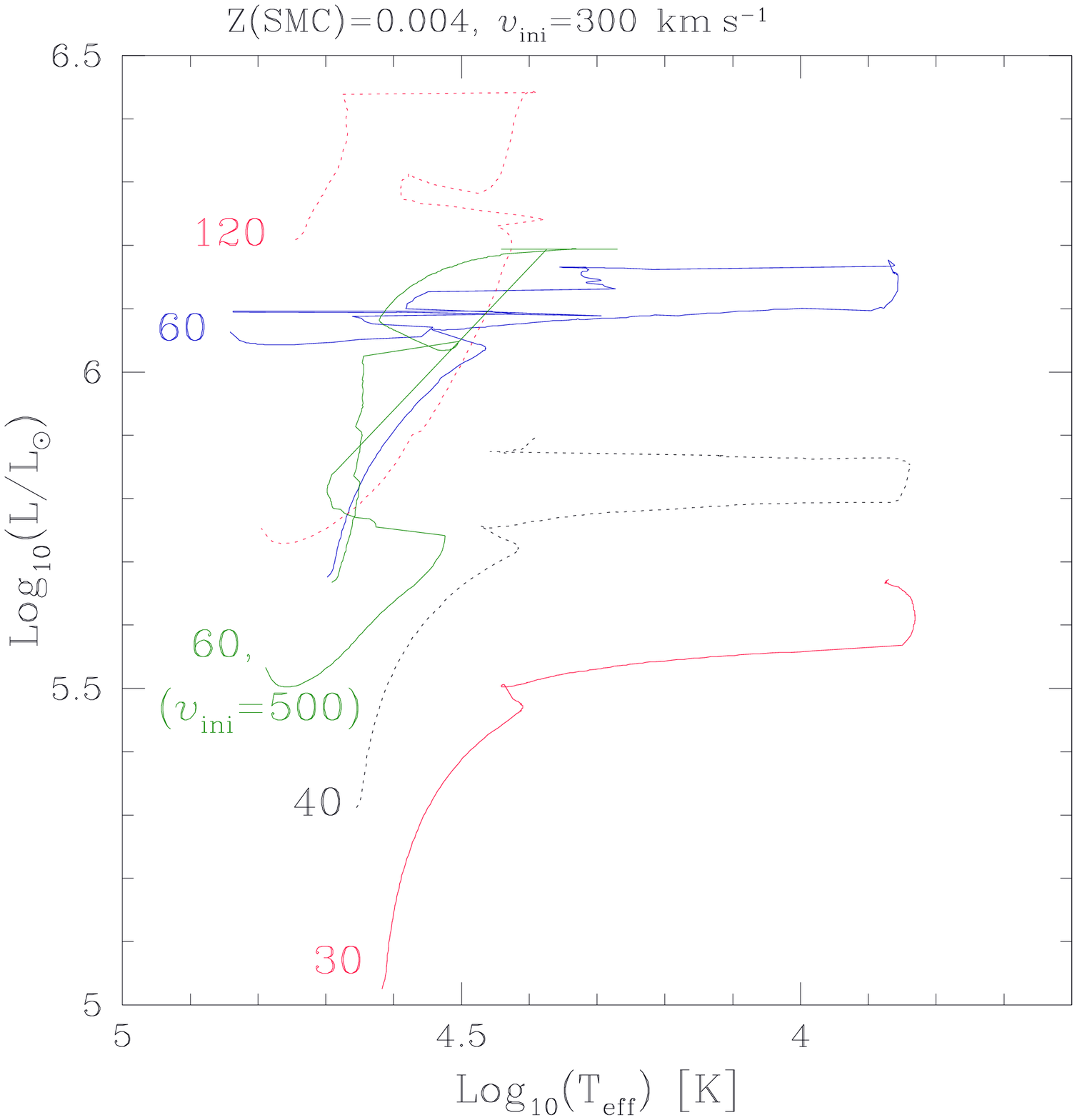}\includegraphics[width=7cm]{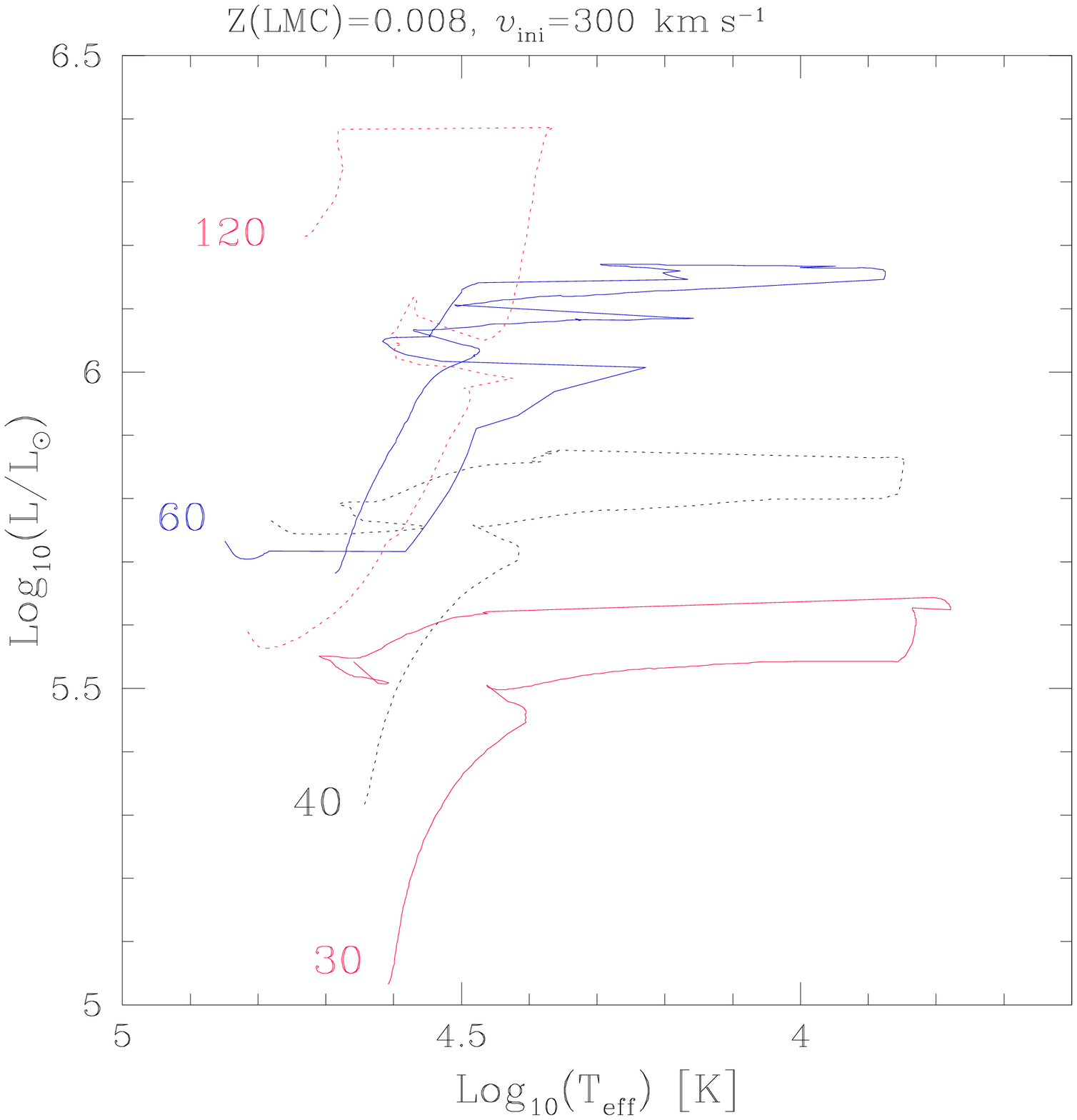}
\caption{Evolution of the $Z=0.004$ ({\it left}) and $Z=0.008$ ({\it right}) 
models in the HR diagram. The evolutionary tracks are drawn using the data available on the web
at http://obswww.unige.ch/$\sim$dessauges/evol/results.html.}
\label{z048}
\end{figure}
\citet{KK04} find a metallicity [Fe/H]$=-0.4 \pm 0.3$ for V838 Mon.
V838 Mon therefore has a sub-solar metallicity close to that of the Large Magellanic
Cloud (LMC: Z $\sim$0.008). The lower bound for metallicity is close to that of the Small 
Magellanic Cloud (SMC: Z $\sim$0.004 or [Fe/H]$\sim$-0.7).
We therefore concentrate on LMC (Z=0.008) metallicity models to find a progenitor,
while solar (Z=0.020) and SMC (Z=0.004) metallicity models are representative for models at
the upper and lower bounds of the metallicity determination for V838 Mon.

How does the evolution in the HR diagram changes between solar and LMC
metallicities? The opacities being lower, the models are a bit bluer and hotter 
during the main sequence. The main difference comes however from mass loss. 
This means that the models below 20 $M_\odot$ are not significantly different 
between solar and LMC metallicities. 
The evolution of the sub-solar metallicity models in the HR diagram is 
shown in Fig. \ref{z048} for models more massive than 20 $M_\odot$.
The minimum initial mass model, which forms a WR star increases to 25 
$M_\odot$ at $Z_{\rm LMC}$ (and 32 $M_\odot$ at $Z_{\rm SMC}$). 
The minimum mass model to evolve 
directly to the WR stage without going through the RSG stage increases as well.
For example, the 60 $M_\odot$ models at $Z_{\rm LMC}$ (and $Z_{\rm SMC}$) 
do go through the RSG stage.
Therefore the lifetime 
in the WR stage decreases with metallicity. 
and, as said earlier (except for the very massive models with $M \sim 100\,M_\odot$), 
the final mass is higher at lower metallicity.

Can a WR star be the progenitor of V838 Mon? WR models have effective 
temperatures between 50\,000 and 25\,000 K compatible with observations. 
The most massive ones can lose up to 100 $M_\odot$ during their lifetime
corresponding to the amount of circumstellar material determined by Ashok.
On the other hand, they are very luminous $L \gtrsim 10^5\,L_\odot$ and it
seems hard to reconcile it with a $V=15$ and $B-V=0.5$ progenitor 
at 6 kpc. WR are also unlikely to evolve to the RSG. 
Using models at $Z_{\rm SMC}$
does not help the comparison with the observations of V838 Mon 
progenitor. 

What other possibilities are there? 
A main sequence single star with an initial mass around 18 $M_\odot$ is not a probable 
progenitor because it is still very luminous and because the time it takes to cross the 
HR diagram to the RSG stage is thousands of years. One advantage is nevertheless that the 
such a RSG reaches very low effective temperatures, around 3500 K, which 
is required by present observations.
V838 Mon companion is a B3V star and the progenitor star was not so different 
from the companion \citep{MH05}. It is therefore possible that the progenitor was also 
a main sequence B star \citep{TS06}. 
The 2002 event cannot be due to the 
usual crossing of the HR diagram at the end of the main sequence because a 
normal star takes too much time to become a RSG and its luminosity does not 
change drastically before or during the crossing. 
Can it be related to the Be phenomenon (B star with a disk)? 
Is it due to a mass transfer from a similar mass companion?
These pathways have not been explored yet, so it is difficult to say whether they 
could reproduce the observations or not. 
The other progenitor models are: a merger \citep{TS06}, a star capturing its 
planets \citep{RZSL06}  and a
born-again low mass star \citep{L05}.
The references given for each model discuss in detail their advantages and 
negative points. It would be very interesting to estimate frequencies
of the different models as well as the frequency of objects like V838 Mon 
(including or not M31 RV and V4332 Sgr) and to compare them to gain
additional constraints.

\section{Conclusion}
The evolution of stellar models for massive rotating stars including mass loss 
was presented with an emphasis on the evolution in the Hertzsprung--Russell
diagram. In particular, models of WR stars and of single massive stars were compared to 
the observations of the V838 Mon. Even though there is no simple evolutionary scenario
able to explain V838 Mon eruption, it is interesting to note a few similarities. 
WR stars, although too luminous, are able to lose up to 100 $M_\odot$ of material 
during their lifetime. Massive stars reach lower RSG effective temperatures, down to 
$3500$ K. 

The usual crossing of the HR diagram by a single star after the main 
sequence cannot explain the 2002 eruption due to timescale and luminosity problems. 
The influence of accretion from a binary companion or disk formation should 
be investigated. Additional constraints can be put on the progenitor models by 
determining the origin (interstellar or circumstellar) and the mass of the material producing the 
light echoes and by estimating the frequency of objects like V838 Mon.

\acknowledgements 
I am grateful to the organising committees for their invitation and help.





\end{document}